\documentclass[5p,number,sort&compress,plb]{elsarticle}

\usepackage{threeparttable}
\usepackage{amsmath}
\usepackage{amsfonts}
\usepackage{multirow}
\usepackage[colorlinks]{hyperref}
\usepackage{amssymb}
\DeclareMathAlphabet{\mathpzc}{OT1}{pzc}{m}{it}
\DeclareMathOperator{\cosec}{cosec}
\usepackage{float}
\usepackage{longtable}
\usepackage{graphicx}

\begin{document}

\begin{frontmatter}	
\title{Determination of $1p$ and $2p$ stripping excitation functions for
$^{16}$O+$^{142}$Ce using a Recoil Mass Spectrometer}
	
\author[iuac]{Rohan Biswas}
\author[iuac]{S. Nath\corref{cor1}}
\ead{subir@iuac.res.in}
\author[iuac]{J. Gehlot}
\author[iuac]{Gonika}
\author[iuac]{Chandra Kumar} 
\author[du] {A. Parihari}
\author[iuac]{N. Madhavan}
\author[ku]{A. Vinayak}
\author[cuj]{Amritraj Mahato}
\author[thapar]{Shoaib Noor}
\author[du]{Phurba Sherpa}
\author[titech,tsukuba]{Kazuyuki Sekizawa}
\ead{sekizawa@phys.titech.ac.jp}

\cortext[cor1]{Corresponding author}
\address[iuac]{Nuclear Physics Group, Inter-University Accelerator Centre,
Aruna Asaf Ali Marg, New Delhi 110067, India}
\address[du]{Department of Physics and Astrophysics, Delhi University,
Delhi 110007, India}
\address[ku]{Department of Physics, Karnatak University,
Dharwad 580003, India}
\address[cuj]{Department of Physics, Central University of Jharkhand,
Ranchi 835205, India}
\address[thapar]{School of Physics and Materials Science, Thapar Institute of
Engineering and Technology, Patiala 147004, India}
\address[titech]{Department of Physics, School of Science,
Tokyo Institute of Technology, Tokyo 152-8551, Japan}
\address[tsukuba]{Nuclear Physics Division, Center for Computational Sciences,
University of Tsukuba, Ibaraki 305-8577, Japan}
	
\date{\today}

\begin{abstract}
We report the first direct measurement of differential transfer cross sections 
using a Recoil Mass Spectrometer. Absolute differential $1p$ and $2p$-stripping
cross sections at $\theta_\mathrm{c.m.}=180^\circ$ have been determined for the
system $^{16}$O+$^{142}$Ce by detecting the heavier target-like ions at the focal
plane of the Heavy Ion Reaction Analyzer. Focal plane spectra have been compared
with the results of a semi-microscopic Monte-Carlo simulation to unambiguously
identify the transfer channels. Transmission efficiency of the target-like ions
through the spectrometer has also been estimated using the simulation which has
been crucial to extract the cross sections from the yields of ions measured 
during the experiment. The methodology adopted in this work can be applied to
measure multi-nucleon transfer cross sections using other similar recoil
separators. The experimental excitation functions for the reactions
$^{142}\mathrm{Ce(}^{16}\mathrm{O,}^{15}\mathrm{N)}^{143}\mathrm{Pr}$ and
$^{142}\mathrm{Ce(}^{16}\mathrm{O,}^{14}\mathrm{C)}^{144}\mathrm{Nd}$
have been compared with coupled reaction channel calculations. An excellent
matching between measurement and theory has been obtained. For $1p$-stripping,
major contribution to the cross section has been found to be the transfer of a
proton from $^{16}\mathrm{O}$ to the $2d_{\frac{5}{2}}$ excited state of
$^{143}\mathrm{Pr}$, leaving behind $^{15}\mathrm{N}$ in the $1p_{\frac{1}{2}}$
ground state. Transfer of a cluster of two protons from $^{16}\mathrm{O}$ to the
$2^{+}$ excited state of $^{144}\mathrm{Nd}$, resulting in $^{14}\mathrm{C}$ in
the $0^{+}$ ground state, appears to be the most probable cause for $2p$-stripping.
Measured transfer probabilities for $1p$ and $2p$ channels have been compared
with Time-Dependent Hartree-Fock calculations. Proton stripping channels are found
to be more favourable compared to neutron pick-up channels. However, the theory
overpredicts the measurement hinting at the need for extended approaches with
explicit treatment of pairing correlations in the calculations.

\end{abstract}

\end{frontmatter}

\section{Introduction}
The simplest picture of a nuclear reaction is a light projectile ion being
scattered off a heavier target nucleus. Such a collision, termed as
\textit{direct nuclear reaction} is characterized by a very short interaction
time $\sim 10^{-22}$ s and active participation of a few nucleons. In contrast,
a \textit{compound nuclear reaction} is much slower in which all the constituent
nucleons of the collision partners take part to form a mono-nucleus. The
resulting `compound nucleus' decays by emission of photons and evaporation of
light particles. Availability of heavy ion beams made a third and intermediate
class of nuclear reactions possible \cite{Kaufmann1961}. Such reactions are
characterized by creation of binary products in the exit channels with broad
mass, charge and angular distributions. Heavy ion-induced reactions, with
features intermediate between direct reactions and compound nuclear reactions,
had been variously termed as deep inelastic collision (DIC), multi-nucleon
transfer (MNT), deep-inelastic transfer, quasi-fission, strongly-damped
collision and relaxation phenomena by different research groups
\cite{Volkov1978}.

MNT reactions have been useful for synthesizing nuclides away from the valley
of $\beta$-stability \cite{Adamian2020}. A renewed interest in the study of
MNT reactions have been ignited by favourable predictions of synthesis of
neutron-rich isotopes of heavy elements \cite{Zagrebaev2008,Adamian2010} in
the recent past. Knowledge about properties of the nuclides in the ‘north-east’
corner of the nuclear chart, in the vicinity of $N = 126$ shell, is very
important for better understanding of stellar nucleosynthesis via the
$r$-process. Transfer of nucleons is also known to influence the dynamics of
fusion between two heavy nuclei \cite{Morton1994,Pollarolo2008}.

Products of MNT reactions had earlier been identified by chemical
separation and measurement of characteristic $\gamma$-rays \cite{Rehm1991}.
Magnetic spectrographs of varied configurations \cite{Enge1979} had also been
used to detect the scattered projectile-like ions from MNT reactions. In the
last two decades, a new class of magnetic separators
\cite{Cappuzzello2016,Rejmund2011,Latina2004} with large acceptance, has been
put to use in the study of MNT reactions
\cite{Ferreira2021,Watanabe2015,Galtarossa2018}. Production of neutron-rich
nuclides in damped collisions has been studied recently with novel applications
of a few recoil separators which were not originally designed and built for this
purpose \cite{Stefan2018,Nitto2018,Devaraja2020}. Production of nuclei far from
the stability region in multi-nucleon transfer reactions using a high resolution
magnetic spectrometer has also been reported recently \cite{Azhibekov2020}.

In MNT reactions, either the projectile-like or the target-like ions can be
detected at the focal plane of the recoil separator \cite{Corradi2009}. In most
separators, usually the lighter projectile-like ions are detected in the forward
angles in the laboratory frame of reference. Mass and charge of the corresponding
target-like ions can be deciphered from two-body collision kinematics. Conversely,
the heavier target-like ions can also be detected at the forward angles. This
technique was first successfully used in the measurement of sub-barrier transfer
reactions for $^{58}$Ni+$^{A}$Sn \cite{Betts1987,Pass1989} using the Daresbury
recoil mass separator \cite{James1988}. Target-like nuclei, separated according to
their mass ($A$) to charge state ($q$) ratio, $\left(\frac{A}{q}\right)$, were
detected by a position-sensitive detector at the focal plane of the separator.
Rehm \cite{Rehm1991} pointed out that such a device had very limited dynamic range
in velocity and charge acceptances. However, the excellent mass resolution
made them quite useful in studying MNT reactions. Similar method was adopted for
measurement of MNT probability, especially at sub-barrier energies, using
other recoil mass spectrometers (RMSs) \cite{Herman1988,Roberts1993,Napoli1993}.
The Heavy Ion Reaction Analyzer (HIRA) \cite{Sinha1994}, the first generation RMS
at IUAC, New Delhi had also been employed to measure few nucleon transfer
probabilities in several medium-heavy systems
\cite{Kataria1997,Baby1997,Sinha1997,Varier1999,Tripathi1999,Kalkal2011}.
All these measurements suffered from two major drawbacks. Firstly, while 
estimating the transfer probability, it had been assumed that the elastic channel
and all the transfer channels had same efficiency for transmission to the focal
plane detector. Secondly, differential transfer cross sections had not been
extracted from the data in most cases, as the transmission efficiency ($\epsilon$)
of the RMS \cite{Nath2007} had not been known. Differential $1n$- and
$2n$-transfer cross sections had been extracted for a limited number of reactions
\cite{Betts1987,Pass1989,Herman1988} with the additional assumption that the sum
of differential elastic, inelastic and transfer cross sections was equal to the
differential Rutherford scattering cross section at energies near and below the
Coulomb barrier. Biswas \textit{et al.} recently reported a methodology
\cite{Biswas2020,Biswas2021} to overcome these assumptions and \textit{measure}
differential quasi-elastic scattering cross sections in an RMS.

In this Letter, we report the first direct measurement of differential transfer
cross sections using an RMS. Differential cross sections for the reactions
$^{142}\mathrm{Ce(}^{16}\mathrm{O,}^{15}\mathrm{N)}^{143}\mathrm{Pr}$ and
$^{142}\mathrm{Ce(}^{16}\mathrm{O,}^{14}\mathrm{C)}^{144}\mathrm{Nd}$ have been
extracted from yields recorded in the experiment. Details of the experiment are
narrated in Section \ref{ExptDetails}. We present results of the experiment in
Section \ref{ExptResults}. Coupled Reaction Channel (CRC) and Time-Dependent
Hartree-Fock (TDHF) calculations, are described in Sections \ref{CRCResults}
and \ref{TDHFResults}, respectively. Finally, in Section \ref{Conclude}, we
summarize and conclude our work.

\section{The experiment}
\label{ExptDetails}
The experiment has been performed in two runs with a pulsed beam of $^{16}$O
ions, having a pulse separation of 4 $\mu$s, accelerated through the 15UD
Pelletron accelerator at IUAC, New Delhi. Isotopically enriched $^{142}$Ce
target foils of thickness $\sim$121.7 $\mu$g/cm$^2$, sandwiched between two
layers ($\sim$20 $\mu$g/cm$^2$ backing and $\sim$5 $\mu$g/cm$^2$ capping) of
graphite films have been used as the target \cite{RBVac2021}. The targets
also contained $\sim 8\%$ impurity of $^{140}$Ce, which had been verified
experimentally \cite{RBVac2021}. Beam energy ($E_{\textrm{lab}}$) has been 
varied between 57 \textendash 69 MeV. Two solid state silicon detectors 
(SSSD), each with a circular aperture of 1 mm diameter, have been placed at
$\theta_{\textrm{lab}} = 15^\circ$ in the horizontal plane at a distance of
100 mm from the target. These detectors have been used as beam monitors
during the experiment and for normalization of cross-sections. The HIRA has
been kept at $\theta_{\textrm{lab}} = 0^\circ$ with an opening aperture of 5
mSr, corresponding to an angular acceptance of $2.2^\circ$. A thin ($\sim 10$
$\mu$g/cm$^2$) graphite foil has been placed 10 cm downstream from the target
to reset charge states of the reaction products to equilibrium distribution.
A Multi-Wire Proportional Counter (MWPC) of dimensions 150 mm in $\mathpzc{x}$
and 50 mm in $\mathpzc{y}$ has been placed at the focal plane of the HIRA.
Here, $\mathpzc{x-z}$ and $\mathpzc{y-z}$ denote the dispersive and the
non-dispersive planes of the RMS, respectively, with $\mathpzc{z}$ being 
direction of the beam. Target-like ions, originating from quasi-elastic
reactions, have been detected at the focal plane of the HIRA, spatially
separated according to their $\frac{\mathpzc{A}}{\mathpzc{q}}$. Energy loss 
$(\Delta E)$ information has been obtained from the cathode of the MWPC. A
Time-to-Amplitude Converter (TAC) has been set up to measure time-of-flight
(TOF) of the ions through the HIRA, in which the anode signal from the MWPC
and the radio-frequency (0.25 MHz RF) signal used for beam pulsing have been
the start and the stop pulses, respectively.

\begin{figure}[ht!]
	\includegraphics[width=\linewidth]{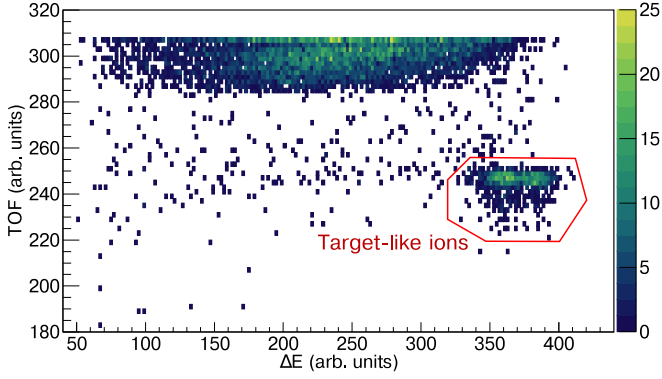}
	\caption{\label{fig:detof} Scatter plot between $\Delta E$ and TOF of the
	events recorded at the focal plane of the HIRA for
	$^{16}\mathrm{O+}^{142}\mathrm{Ce}$ at $E_\mathrm{lab}=63$ MeV. The 
	target-like ions are marked.}
\end{figure}

Fig. \ref{fig:detof} shows the $\Delta E$-TOF spectrum of the events detected
at the focal plane of the HIRA at $E_\textrm{lab}=63$ MeV. Target-like ions
have been identified by their higher $\Delta E$ and closely-related TOF.
The $\mathpzc{x}$-TOF spectrum at the same $E_\textrm{lab}$, gated with these
events, are shown in Fig. \ref{fig:xtof}(a). The most intense group with
$\frac{\mathpzc{A}}{\mathpzc{q}} = \frac{142}{14}$ corresponds to elastically
/ inelastically scattered $^{142}$Ce ions. Another charge state group of
target-like nuclei from the elastic / inelastic channel(s), marked with
$\frac{142}{15}$, can be seen in the extreme left of the figure. It is
important to note here that we can not differentiate the inelastic channel(s)
from the elastic channel in our method of measurement. Origin of the group
marked with $\frac{140}{14}$ is ambiguous. These ions may arise from elastic
/ inelastic scattering between the projectile nuclei and nuclei of $^{140}$Ce,
present as impurities in the target foil \cite{RBVac2021}. Probable pickup
of two neutrons from $^{142}$Ce ($Q_{0}^{\mathrm{+2n}} = -0.41$ MeV) may also
lead to formation of $^{140}$Ce. Target-like ions, corresponding to probable
$1n$-pickup ($Q_{0}^{\mathrm{+1n}} = -3.03$ MeV) channel, have not been
identified as a distinct group in the $\mathpzc{x}$-TOF spectrum. The groups
marked with $\frac{143}{14}$ and $\frac{144}{14}$ corresponding to $1p$- 
($Q_{0}^{\mathrm{-1p}} = -6.303$ MeV) and $2p$-stripping
($Q_{0}^{\mathrm{-2p}} = -8.54$ MeV) channels, however, can be distinctly
observed. While the recoiling target-like nuclei from $1p$ channel
($^{143}$Pr), have been recorded over the entire range of projectile energies,
the same from $2p$ channel, ($^{144}$Nd), have been observed only at a few
energies near the barrier, within the limited duration that have been available
for collecting the data. It must be stated here that ascertaining the charge of
the transferred nucleon(s) from the spectrum (Fig. \ref{fig:xtof}(a)) alone is
not possible. Identification of transfer channels have been realized with
recourse to $Q$-value considerations.
 
Some residual background events, originating from multiple-scattering of
projectile ions inside the spectrometer, can be observed in the measured
$\mathpzc{x}$-TOF spectrum. Further rejection of background events can be
achieved by making use of the method of kinematic coincidence between the
recoiling target-like ions detected at the focal plane of the RMS and the
back-scattered projectile-like ions detected by a $\Delta E - E$ telescope
\cite{Sinha1997}.

\section{Results}
\label{ExptResults}
\begin{figure}[ht!]
	\includegraphics[width=\linewidth]{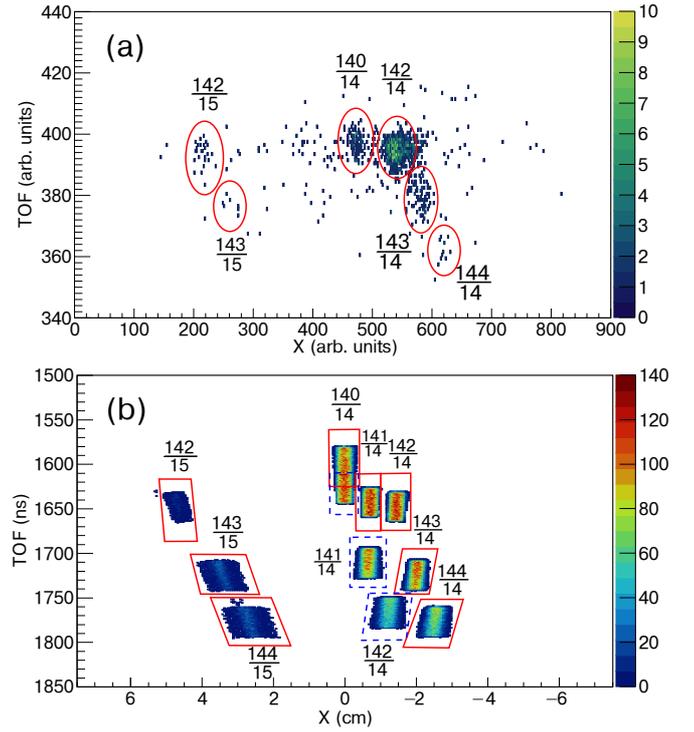}
	\caption{\label{fig:xtof} (a) Experimental and (b) simulated $\mathpzc{x}$-TOF
	spectra of target-like ions for the system $^{16}\mathrm{O+}^{142}\mathrm{Ce}$
	at $E_\mathrm{lab} = 63$ MeV. The identified channels are marked in panel (a).
	In panel (b), position of events from the elastic, $1n$ and $2n$ pick-up and $1p$
	and $2p$ stripping channels for the system $^{16}\mathrm{O+}^{142}\mathrm{Ce}$
	are marked within boxes with solid (red) lines. Position of events from the
	elastic, $1p$ and $2p$ stripping channels for the system
	$^{16}\mathrm{O+}^{140}\mathrm{Ce}$ are also shown in panel (b) and marked
	within boxes with dashed (blue) lines. See text for details.}
\end{figure}

The elastic/inelastic and transfer channels in the experimental spectrum
(Fig. \ref{fig:xtof}(a)) have been identified based on their
$\frac{\mathpzc{A}}{\mathpzc{q}}$ values. To rule out any ambiguity in
channel identification, the experimental spectrum has been further compared
with a simulated $\mathpzc{x}$-TOF spectrum, which is shown in Fig.
\ref{fig:xtof}(b). The simulation has been carried out by a semi-microscopic
Monte Carlo code \cite{Biswas2020,Biswas2021}. Simulated positions of the 
target-like ions, generated for the system $^{16}$O+$^{142}$Ce, \textit{viz.}
$^{142}$Ce, (elastic/inelastic) $^{141}$Ce ($1n$-pickup), $^{140}$Ce
($2n$-pickup), $^{143}$Pr ($1p$-stripping) and $^{144}$Nd ($2p$-stripping),
are marked in Fig. \ref{fig:xtof}(b). In addition, simulated positions of
target-like ions which may result from collisions between $^{16}$O and 
$^{140}$Ce (impurity in the target foil), \textit{viz.} $^{140}$Ce
(elastic/inelastic), $^{141}$Pr ($1p$-stripping) and $^{142}$Nd
($2p$-stripping), are also marked in the figure. Comparing the two panels of
Fig. \ref{fig:xtof}, we may conclude that (i) the target-like ions resulting
from $1n$-pickup in the system $^{16}$O+$^{142}$Ce (marked $\frac{141}{14}$
in the top half of panel (b)) can be distinguished from the 
elastic/inelastic products, if produced in the experiment, (ii) the two
probable source of the group with 
$\frac{\mathpzc{A}}{\mathpzc{q}} = \frac{140}{14}$ can not be resolved in
TOF, (iii) the identification of $1p$- and $2p$-stripping channels for the
system $^{16}$O+$^{142}$Ce is unambiguous and (iv) the target-like ions
resulting from $1p$- and $2p$-stripping in the system $^{16}$O+$^{140}$Ce
(marked $\frac{141}{14}$ and $\frac{142}{14}$ in the bottom half of panel
(b)) should be identifiable, if produced in the experiment.
This comparison underlines the need for highly-enriched isotopic target
foils for such studies with RMS.

The absolute differential cross-sections for $1p$ and $2p$-stripping at
centre of mass (c.m.) angle, $\theta_\mathrm{c.m.} = 180^\circ$ have been
calculated using the relation
\cite{Biswas2021}:

\begin{equation}
		\label{eq:XS_p}
		\left(\frac{d\sigma}{d\Omega}\right)^{\textrm{1p(2p)}}_{180^{\circ}}
		= \frac{Y_{\textrm{143(144)}}}{Y^{\textrm{Ruth}}_{\textrm{norm}}}
		\frac{\Omega_{\textrm{norm}}}{\Omega^{\textrm{eff}}_{\textrm{HIRA}}}
		\frac{1}{\epsilon_{\textrm{HIRA}}}
		\left(\frac{d\sigma}{d\Omega}\right)^{\textrm{Ruth}}_{\theta_{\textrm{norm}}}
\end{equation}

\noindent
where $Y_{143(144)}$ is the yield of the $\frac{143}{\mathpzc{q}}$
$\left(\frac{144}{\mathpzc{q}}\right)$ group(s) in the $\mathpzc{x}$-TOF
spectrum (Fig. \ref{fig:xtof}) corresponding to $1p$-stripping ($2p$-stripping).
$Y^{\textrm{Ruth}}_{\textrm{norm}}$ is the geometric mean of the counts recorded
in the two normalization detectors. $\Omega_{\textrm{norm}}$ and $\left(\frac{d\sigma}{d\Omega}\right)^{\textrm{Ruth}}_{\theta_{\textrm{norm}}}$
are the solid angle subtended by each of the normalization detectors and the
differential Rutherford scattering cross section in the c.m. frame of reference
at $\theta_{\textrm{norm}}=16.32^\circ$ (corresponding to
$\theta_{\textrm{lab}} = 15^\circ$), respectively. The transmission efficiency
$\epsilon_{\textrm{HIRA}}$ for the target-like ions have been calculated
\cite{Biswas2020} by taking the ratio of the counts of ions reaching the focal
plane to the number of ions entering the entrance aperture of the HIRA. The
effective solid angle, $\Omega^\textrm{eff}_\textrm{HIRA}$, has been determined
experimentally by recording the target-like ions at $E_{\textrm{lab}} = 48$ MeV
using the relation \cite{Biswas2021}:

\begin{equation}
	\label{eq_Omeg_RMS_eff}
	\Omega^{\textrm{eff}}_{\textrm{HIRA}} =
	\frac{Y^{\textrm{Ruth}}_{\textrm{142}}}{Y^{\textrm{Ruth}}_{\textrm{norm}}}
	\Omega_{\textrm{norm}}
	\frac{1}{\epsilon_{\textrm{HIRA}}}
	\left(\frac{d\sigma}{d\Omega}\right)^{\textrm{Ruth}}_{\theta_{\textrm{norm}}}
	\ /\ 
	\left(\frac{d\sigma}{d\Omega}\right)^{\textrm{Ruth}}_{180^{\circ}} .
\end{equation} 

\begin{figure}[ht!]
	\centering
	\includegraphics[width=\linewidth]{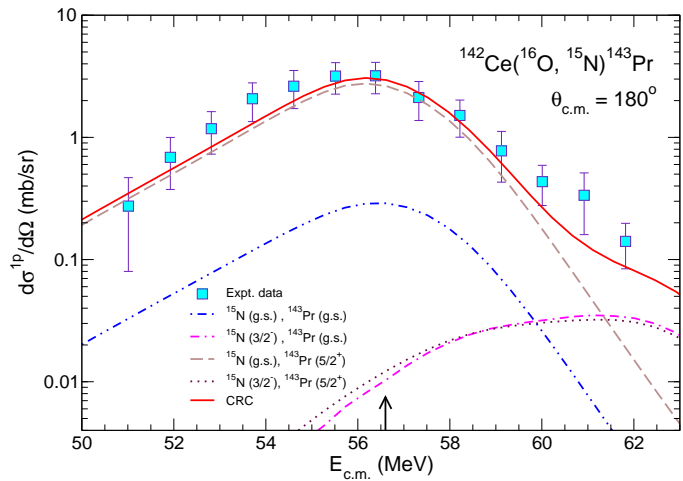}
	\caption{\label{fig:cs1p} Absolute differential cross sections for the reaction
	$^{142}$Ce($^{16}$O,$^{15}$N)$^{143}$Pr at $\theta_{\mathrm{c.m.}} = 180^{\circ}$
	as a function of $E_\mathrm{c.m.}$. The full CRC calculation is shown by the 
	solid (red) line. Contributions of different exit channels to the transfer
	cross sections are also shown. The arrow denotes location of the Coulomb barrier.}
\end{figure}

\begin{figure}[ht!]
	\centering
	\includegraphics[width=\linewidth]{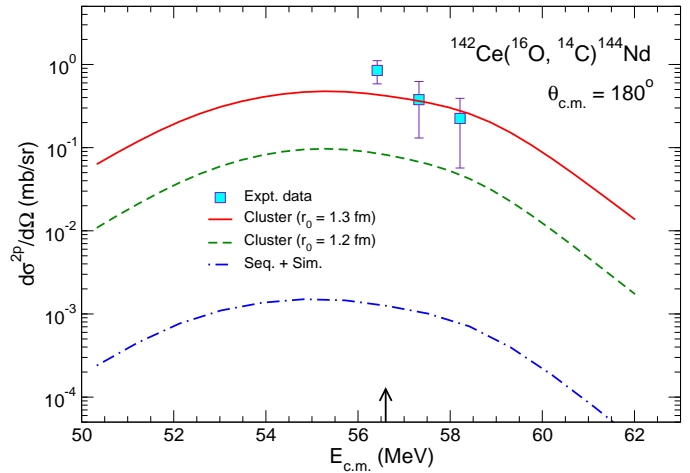}
	\caption{\label{fig:cs2p} Absolute differential cross sections for the
	reaction $^{142}$Ce($^{16}$O,$^{14}$C)$^{144}$Nd
	at $\theta_{\mathrm{c.m.}} = 180^{\circ}$ as a function of $E_\mathrm{c.m.}$.
	Results of the combined sequential and microscopic simultaneous CRC
	calculations are shown by the dot-dashed (blue) line. Results of the extreme
	cluster model CRC calculations are shown by solid (red) line ($r_0=1.3$ fm)
	and dashed (green) lines ($r_0=1.2$ fm). The arrow denotes location of the
	Coulomb barrier.}
\end{figure}

At this energy ($\simeq$ 25\% below the Coulomb barrier), all scattering events
obey Rutherford scattering and the transfer channels are closed. The absolute 
differential cross sections for $1p$ and $2p$ stripping channels, as a function
of the energy available in the c.m. frame of reference, $E_{\textrm{c.m.}}$ are,
respectively, shown in Fig. \ref{fig:cs1p} and Fig. \ref{fig:cs2p}.

Error bars in Fig. \ref{fig:cs1p} and Fig. \ref{fig:cs2p} include statistical as
well as systematic uncertainties, the latter of which are listed in Table 
\ref{tab:sys_err}. The error in $\Omega^{\textrm{eff}}_{\textrm{HIRA}}$ contains 
statistical uncertainties in the measured yields
($Y^{\textrm{Ruth}}_{\textrm{142}}$ and $Y^{\textrm{Ruth}}_{\textrm{norm}}$)
at $E_\mathrm{lab} = 48$ MeV and similar systematic uncertainties. 


\begin{table}[h!]
   \centering
      \begin{threeparttable}
         \caption{\label{tab:sys_err} Systematic errors considered while
         extracting the experimental differential $1p$ and $2p$-stripping
         cross-sections.}
         \begin{tabular}{lcc}
            \hline
            Quantity                   & Uncertainty       & \% effect \\
            \hline
            $\theta_{\textrm{norm}}$   & 0.5$^\circ$       & 12.8 \tnote{a} \\
            \multirow{2}{*}{$\Omega_{\textrm{norm}}$} & 2.0 mm \tnote{b} & \multirow{2}{*}{4.0} \\
                                       & 0.01 mm \tnote{c} & \\
            $\epsilon_{\textrm{HIRA}}$ & 10.0 \%             & 10.0 \\
            $\Omega^\textrm{eff}_\textrm{HIRA}$ & 17.1 \% & 17.1 \\
            \hline  
         \end{tabular}
         
         \begin{tablenotes}
         \begin{footnotesize}
            \item[a]{Error in calculated
            $\left(\frac{d\sigma}{d\Omega}\right)^{\textrm{Ruth}}_{\theta_{\textrm{norm}}}$}
            \item[b]{Uncertainty in distance between target and detector}
            \item[c]{Uncertainty in aperture diameter}
         \end{footnotesize}
         \end{tablenotes}
      \end{threeparttable}
\end{table}

\section{Coupled Reaction Channel calculations}
\label{CRCResults}

\begin{figure}[hb!]
	\centering
	\includegraphics[width=\linewidth]{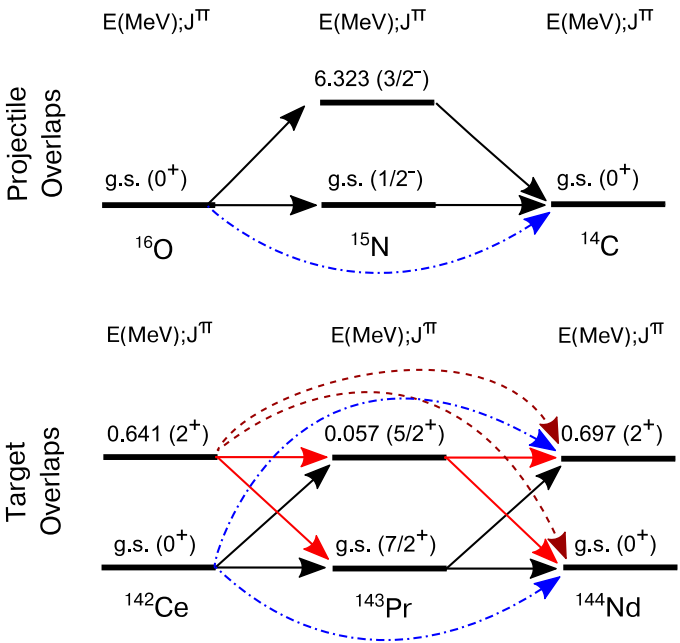}
	\caption{\label{fig:ovl1p} Coupling scheme adopted for CRC calculations of
	$1p$ (solid arrows) and $2p$ (dash-dotted arrows for g.s. of entrance channel
	and dashed arrows for excited state of entrance channel) transfer.}
\end{figure}

Measured differential cross sections have been compared with finite-range
CRC model calculations performed using the code \textsc{fresco}
\cite{fresco_web,fresco_cpr}. The coupling scheme is shown in Fig.
\ref{fig:ovl1p}. Reduced transition probability ($B(E2)$) for transition to the
$2^+$ state of $^{142}$Ce has been taken to be 4572 e$^2$fm$^4$
\cite{Pritychenko2016}. Entrance channel, exit channel and core-core
($^{15}$N+$^{142}$Ce) interactions for $1p$-stripping channel have been defined
by optical potential containing real and imaginary parts, taken in Woods-Saxon
shape where the parameters have been determined using Aky\"{u}z-Winther formalism
\cite{AW}. The parameters of the optical potential, used in the calculations, are
listed in Table \ref{tab:pot}. The real parameters have been determined at
$E_\textrm{lab} = 63$ MeV. Parameters of the imaginary part of the entrance
channel optical potential have been determined by reproducing the
elastic/inelastic scattering excitation function. The Woods-Saxon parameters for
the proton bound-state potential in $^{16}$O have been taken from Ref.
\cite{Leuschner1994}. Global parameters \cite{Koning2003} have been used to
define the proton bound-state potential in $^{143}$Pr. Depth of the real part of
bound-state potentials has been varied to reproduce the binding energy of the 
proton to the core. Spectroscopic factors for the $1p_{\frac{1}{2}}$ ground state
(g.s.) and $1p_{\frac{3}{2}}$ third excited state of $^{15}$N have been taken to
be 1.275 and 2.047, respectively \cite{Leuschner1994}. Spectroscopic factors for
the overlaps between states in the target-like nuclei have been taken to be
$1.0$. The calculations have been performed in prior representation including
full complex remnant terms.

\begin{table*}[ht!]
	\centering
	\caption{\label{tab:pot} Optical potential parameters used to define 
	interaction between different channels. Depth ($V_0$) of the real part of
	the potential is taken at $E_\mathrm{lab} = 63$ MeV.}
	\begin{tabular}{ccccccc}
		\hline
		System & V$_0$ (MeV) & r$_0$ (fm) & a$_0$ (fm) & W$_0$ (MeV) & r$_w$ (fm) & a$_w$ (fm)    \\
		\hline
		$^{16}$O+$^{142}$Ce & 61.859 & 1.2 & 0.651 & 50.0   & 1.0 & 0.4   \\
		$^{15}$N+$^{143}$Pr & 59.571 & 1.2 & 0.647 & 14.893 & 1.0 & 0.4   \\
		$^{15}$N+$^{142}$Ce & 59.387 & 1.2 & 0.649 & 14.847 & 1.2 & 0.649 \\
		$^{14}$C+$^{144}$Nd & 57.070 & 1.2 & 0.647 & 14.267 & 1.2 & 0.647 \\
		$^{14}$C+$^{143}$Pr & 56.893 & 1.2 & 0.647 & 14.223 & 1.2 & 0.647 \\
		$^{14}$C+$^{142}$Ce & 56.714 & 1.2 & 0.647 & 14.178 & 1.2 & 0.647 \\
		\hline
	\end{tabular}
\end{table*}

Results of CRC calculations agree very well with the experimental cross sections
for $1p$ stripping, as shown in Fig. \ref{fig:cs1p}. The major contribution to the
cross-sections is by the proton transfer to the $2d_{\frac{5}{2}}$ state of 
$^{143}$Pr and g.s. to g.s. transfer between $^{16}$O and $^{15}$N. The hole state
$1p_{\frac{3}{2}}$ in $^{15}$N, which is at much higher excitation energy than the
target-like nuclei, contribute to the cross sections at higher projectile energies. 

To understand the mechanism for $2p$ transfer in the reaction
$^{142}$Ce($^{16}$O,$^{14}$C)$^{144}$Nd,
CRC calculations have been performed considering (a) combination of sequential
(two-step) and microscopic simultaneous (one-step) processes and (b) extreme cluster
model. Couplings considered in the sequential and microscopic simultaneous processes
are shown in Fig. \ref{fig:ovl1p}. The excited states in the intermediate nuclei have
been taken to be the same as that for $1p$ transfer. Global optical model parameters
for the $^{14}$C+$^{144}$Nd exit channel and $^{14}$C+$^{143}$Pr core-core interaction
are given in Table \ref{tab:pot}. The one-proton binding potential has been determined 
using the global optical model potential \cite{Koning2003} where the depth of the real
part has been adjusted to reproduce the one-proton binding energies to the $^{14}$C and
$^{143}$Pr cores. Spectroscopic amplitudes of 0.9141 and 0.2867, respectively, have
been used for the overlaps
$\left\langle ^{15}\mathrm{N}(\frac{1}{2}^-)\middle|^{14}\mathrm{C}(0^+)\right\rangle$
and $\left\langle ^{15}\mathrm{N}(\frac{3}{2}^-)\middle|^{14}\mathrm{C}(0^+)\right\rangle$
\cite{Tung1978}. Spectroscopic amplitudes for overlaps between $^{143}$Pr and $^{144}$Nd
have been considered to be 1. The calculations have been done in prior-post combination
to avoid the non-orthogonality terms. The same interaction potentials have been used in
the microscopic simultaneous transfer process. The ground state ($\psi_\mathrm{g.s}$) of
$^{16}$O has been taken as \cite{Tung1978}:

\begin{equation*}
	\label{eq:g.s.}
	\psi_\mathrm{g.s.}= 0.914\left(1p_{\frac{1}{2}}\right)^2 +
	0.405\left(1p_{\frac{3}{2}}\right)^2
\end{equation*}

In this case, depth of the binding potential has been adjusted to reproduce half the
$2p$ separation energies. The calculations have been performed in prior form. In both
the above methods, full complex remnant terms have been included in the calculations.
The combined calculation underpredicts the experimental data by three orders of 
magnitude, as shown in Fig. \ref{fig:cs2p}. There is a minor increase in the cross 
sections (not shown in the figure) when the radius of binding potential well is 
changed to 1.3 fm. However, the increase remains insignificant with respect to the
data.

In the extreme cluster model \cite{Satchler} analysis, the $2p$ pair is considered
to be anti-symmetric to each other with $S = 0$ and $T = 0$. In the harmonic
oscillator potential, the principle quantum number and orbital angular momentum
parameters corresponding to individual protons ($n_\mathrm{i}$, $\ell_\mathrm{i}$;
i = 1, 2) are transformed into ($n$, $\ell$) and ($N$, $L$), corresponding to
parameters relative to each other and to the c.m. of the core-cluster system,
respectively. They are related through energy conservation by \cite{Mohinsky1959}:

\begin{equation}
	\label{eq:NL}
	\sum_{\mathrm{i} = 1, 2} 2n_\mathrm{i}+\ell_\mathrm{i} = 2N+L+2n+\ell \ \ .
\end{equation}

Assuming the proton pair to be in $1s$ state ($n=1$, $\ell=0$) relative to each
other, we can find the combination of ($N$,$L$) using Eq. \eqref{eq:NL} to
define the transfer of angular momentum according to the overlaps taken in
Fig. \ref{fig:ovl1p}. The CRC calculations have been performed by assigning
($N=2$, $L=0$) for $\left\langle 0^+\middle|0^+\right\rangle$ transfer
between $^{16}$O and $^{14}$C. For transfer between $^{142}$Ce and
$^{144}$Nd, CRC calculations have been carried out for ($N=5$, $L=0$) for
$\left\langle 0^+\middle|0^+\right\rangle$, 
$\left\langle 2^+\middle|2^+\right\rangle$ transfers and ($N=4$, $L=2$) for
$\left\langle 0^+\middle|2^+\right\rangle$,
$\left\langle 2^+\middle|0^+\right\rangle$ transfers. The two-proton
Woods-Saxon binding potential well has been defined by radius $r_0 = 1.2$ 
fm and diffuseness parameter $a = 0.6$ fm \cite{Jha2002}, while depth of the 
potential has been adjusted to reproduce the $2p$ separation energies.
Spectroscopic amplitudes have been taken to be 1. The calculations have been
performed in prior-form in which full complex remnant terms have been
included. As observed in Fig. \ref{fig:cs2p}, the shape of the excitation
function is the same as that for the sequential plus microscopic simultaneous
calculations but the magnitude is larger by two orders. Yet, the calculations
underpredict the data by an order. As the magnitude depends on the shape of
the binding potential well \cite{Jha2002}, increasing the radius parameter to
$r_0 = 1.3$ fm has resulted in enhancement of the theoretical cross-sections
by an order thus reproducing the experimental excitation function. 

No arbitrary normalization has been included in any of the calculated
excitation functions.

\section{Time-Dependent Hartree-Fock calculations}
\label{TDHFResults}
We have also analyzed the $^{16}$O+$^{142}$Ce reaction based on the
microscopic framework of the Time-Dependent Hartree-Fock (TDHF) theory.
A three-dimensional (3D) parallel TDHF code has been used, which has been
continuously developed and applied for a variety of systems (see Ref.
\cite{Sekizawa(2019)} and references therein), including various extensions
going beyond the standard TDHF approach
\cite{Williams(2018),KS_KH_TDHF+Langevin,KS_AS_Ni-Pb,AS_KS_Xe-Pb}.
Here we provide information relevant to the present analysis. (For details of
the theoretical framework and various applications, see, e.g., recent reviews
\cite{TDHF-review(2018),Sekizawa(2019),Stevenson(2019)}.)

For the energy density functional (EDF), Skyrme SLy4d parameter set \cite{SLy4d}
has been used, which does not involve the c.m. correction in its fitting
protocol \cite{Chabanat(1995)}. Single-particle wave functions are represented
by discretizing 3D Cartesian coordinates into a uniform grid with 0.8 fm grid
spacing. For static Hartree-Fock calculations, a box of 24$^3$\,fm$^3$ has been
used, while a box of 64\,$\times$\,32\,$\times$\,24\,fm$^3$ has been used for
time-dependent simulations. The ground states of doubly-magic $^{16}$O is of
spherical shape. Since the number of neutrons in $^{142}_{58}$Ce$_{84}$ is
close to the $N = 82$ magic number, we have found that it is nearly of spherical
shape in its Hartree-Fock ground state (with a tiny octupole deformation). We set
the incident direction and the impact parameter ($b$) vector parallel to the $-x$
and $+y$ directions, respectively, assigning $x-y$ plane as the reaction plane.
Since the deformation of the target nucleus ($\beta_{2} = 0.1259$) is small, we
consider a single initial orientation of $^{142}$Ce, where a non-axial quadrupole
moment, $Q_{22}\propto \bigl<x^2-y^2\bigr>$, takes the smallest value in the
reaction plane.

Once the EDF is fixed, the TDHF approach does not have adjustable parameters
on reaction dynamics. In this sense, TDHF provides a non-empirical description
of low-energy heavy-ion reactions. However, it is of course not a perfect
framework since, e.g., it misses pairing correlations and mean-field
fluctuations. Apart from the uncertainty inherent in the choice of an EDF,
disagreements between TDHF and measurements could indicate importance of the
physics beyond the TDHF approach. Keeping these points in mind, we compare the
TDHF results with the experimental data for the $^{16}$O+$^{142}$Ce reaction.

Since the measurement has been carried out at $\theta_{\rm c.m.}$\,=\,180$^\circ$,
TDHF calculations have been performed for head-on collisions (\textit{i.e.},
$b$\,=\,0) between $^{16}$O and $^{142}$Ce, with changing collision energies. One
should note that the relative motion of colliding nuclei (mean fields) is
classical in TDHF. Hence, one can observe either fusion or non-fusion (not a
superposition of them) depending on the initial conditions. For the present
reaction at $b = 0$, we have found that $E_{\rm c.m.} \le 56.6$ MeV results in
binary reactions, whereas fusion takes place for $E_{\rm c.m.} \ge 56.7$ MeV.

\begin{figure}[ht]
\begin{center}
\includegraphics[width=0.95\columnwidth]{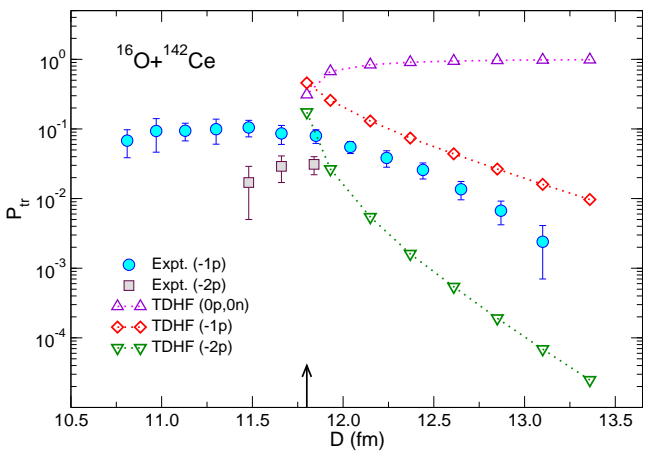}
\caption{
Transfer probabilities in the $^{16}$O+$^{142}$Ce reaction as a function of
$D$. Experimental data for $1p$ and $2p$ stripping channels are shown by solid
symbols. Results of TDHF calculations for quasielastic (0p,\,0n), ($-1$p), and
($-2$p) channels are shown by open symbols, connected with dotted lines.}
\label{FIG:Ptr_16O+142Ce}
\end{center}
\end{figure}

For the binary reactions, we find that transfer of protons is more favorable than
that of neutrons, although the absolute values are small. For instance, the average
number of transferred protons reaches about 0.86 at the maximum for
$E_{\rm c.m.} = 56.6$ MeV, while that of neutrons is rather small, less than 0.06
at the maximum. From a TDHF wave function after collision, one can extract transfer
probabilities using the particle-number projection technique \cite{PNP}. The
extracted probabilities for quasi-elastic [($0p$,\,$0n$), without nucleon transfer],
one-proton stripping ($-1p$) and two-proton stripping ($-2p$) reactions are shown
in Fig. \ref{FIG:Ptr_16O+142Ce} as a function of the distance of the closest
approach. The same is defined as
\begin{equation}
\label{DisCloApp}
D = \frac{Z_{\textrm{p}} Z_{\textrm{t}} e^{2}}{2 E_{\textrm{c.m.}}}
\left(1 + \cosec \frac{\theta_{\textrm{c.m.}}}{2} \right)\ ,
\end{equation}

\noindent
where $Z_{\textrm{p}}$ and $Z_{\textrm{t}}$ are the atomic number of the projectile
and the target, respectively, $\theta_{\textrm{c.m.}}$ is the angle of the
projectile-like ions in the c.m. frame of reference and $e^{2} = 1.44$ MeV fm.
We note that one can obtain $D$ from TDHF time evolution, which gives smaller values
especially close to the fusion threshold. However, here we use Eq.\ref{DisCloApp} for
comparison with the experimental data.

As is apparent from Fig. \ref{FIG:Ptr_16O+142Ce}, processes are dominated
by the quasi-elastic scattering without nucleon transfer \textit{i.e.}
($0p$,\,$0n$) in the sub-barrier regime. Because of quantum tunneling of
single-particle wave functions, there are small, yet finite probabilities for
$1p$ and $2p$ stripping processes. The transfer probabilities increase with
increasing $E_{\textrm{c.m.}}$, since it in turn decreases $D$. At energy close
to the fusion threshold, probabilities of multi-nucleon transfer increase in
TDHF as the system develop neck structure, although such behavior is not seen in
the experimental data at $E_{\rm c.m.}$\,$\simeq$\,56.7\,MeV ($D \simeq 11.8$ fm).
From the figure, we find that TDHF systematically overestimates the transfer
probability for $1p$ stripping (about three times larger) as compared to the
measurements. The experimental data indicates that channels accompanying transfers
of more than one proton are more probable than the TDHF prediction. However,
experimental data for multi-proton transfer are not available in the sub-barrier
regime. It is worth mentioning here that total kinetic energy loss (TKEL) is found
to be at most 7 MeV for $E_{\rm c.m.}$\,=\,56.6 MeV within the TDHF approach.
Thus, particle evaporation effects are expected to be negligible in the
sub-barrier region. It would be interesting to re-examine this reaction
employing extended approaches that treat explicitly pairing correlations
\cite{Ebata(2010),Scamps(2013),Magierski(2017),Hashimoto(2016),Scamps(2017),Regnier(2018)}.
We note that the above-mentioned observation is consistent with the CRC analysis,
\textit{i.e.}, there is a substantial contribution from simultaneous transfer of
two protons in the $^{16}$O+$^{142}$Ce reaction.

\section{Conclusions}
\label{Conclude}
We demonstrate a novel method to measure multi-nucleon
differential transfer cross sections directly, using an RMS, for the first
time. Excitation functions for the reactions 
$^{142}\mathrm{Ce(}^{16}\mathrm{O,}^{15}\mathrm{N)}^{143}\mathrm{Pr}$ and
$^{142}\mathrm{Ce(}^{16}\mathrm{O,}$ $^{14}\mathrm{C)}^{144}\mathrm{Nd}$ have
been measured around the Coulomb barrier. The heavier target-like ions have
been detected at the focal plane of the HIRA, where ions from different
exit channels are dispersed according to their $\frac{A}{q}$ values.
Information of $\Delta E$ and TOF has further helped to reduce the background
caused by randomly-scattered projectile-like ions. Ion trajectories inside
the RMS have been simulated by a Monte-Carlo code to calculate the
transmission efficiency of the target-like ions. The channels have been
unambiguously identified with the aid of a comparison between the measured
and simulated $\mathpzc{x}$-TOF spectra. This methodology can be adopted
for measuring differential quasi-elastic cross sections in other similar
recoil separators. CRC calculations have been performed to understand the
mechanism of $1p$ and $2p$ stripping. Transfer of a proton from $^{16}$O
to the $2d_{\frac{5}{2}}$ excited state of $^{143}$Pr, while $^{15}$N is
left in the g.s., largely contributes to $1p$-stripping cross sections. At
higher excitation energies, contribution from the $1p_{\frac{3}{2}}$ hole
state of $^{15}$N is found to be significant. Transfer of a cluster of two
protons from the g.s. of $^{16}$O to the $2^{+}$ excited state of $^{144}$Nd,
with $^{14}$C remaining in the $0^{+}$ g.s., best reproduces the 
$2p$-stripping cross sections. This observation is similar to other studies
involving two-nucleon transfer \cite{Jha2002,Cavallaro2013}. TDHF
calculations indicate that proton(s) transfer is favoured compared to
transfer of neutron(s) in the present reaction. We have found that TDHF
calculations overpredict measured transfer probabilities, indicating that
simultaneous transfer of two protons has a significant contribution. For
better understanding, it is necessary to use extended approaches in which
pairing correlations are explicitly taken into account.

\vspace{5mm}
\noindent
{\bf{Acknowledgements}}\\
R.B. acknowledges Council of Scientific and Industrial Research (CSIR), New
Delhi for financial support via grant no. CSIR/09/760(0030)/2017-EMR-I. K.S.
used computational resources of the HPCI system (Oakforest-PACS) provided by
Joint Center for Advanced High Performance Computing (JCAHPC) through the HPCI
System Project (Project ID: hp210023) and computational resources (in art) of
the Yukawa-21 System at Yukawa Institute for Theoretical Physics (YITP), Kyoto
University. K.S. was supported by the Japan Society for the Promotion of
Science (JSPS) KAKENHI, Grant-in-Aid for Early-Career Scientists via grant no.
19K14704. The authors are grateful to the Pelletron staff of IUAC for excellent
operation of the accelerator during the experiment and the Target Laboratory
personnel of IUAC for fabrication of target foils. Discussions with Dr. Md.
Moin Shaikh are thankfully acknowledged.

\vspace{10mm}
\noindent
{\bf{References}}

\end{document}